\documentclass[a4paper]{article}
\usepackage{moriond,epsfig} 

\makeatletter
\def\PhAddress#1{\expandafter\def\expandafter\@aabuffer\expandafter
{\@aabuffer \small\it #1\relax 
\begin{figure}[h]
\begin{center}
\epsfig{file=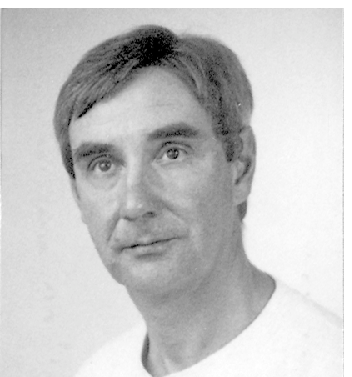}
\end{center}
\end{figure}
\par\vspace{1em}}}
\makeatother

\let\address\PhAddress

\begin{document}
\author{Wojciech S\l omi\'nski and Jerzy Szwed}
\address{Institute of Computer Science, Jagellonian University,
\\ 
Reymonta 4, 30-059 Krak\'ow, Poland
}
\title{REMARKS ON THE
ELECTRON STRUCTURE FUNCTION\footnote{Work supported by
the Polish State Committee for Scientific Research
(grant No. 2~P03B~061~16). Presented by Jerzy Szwed}}
\maketitle
\begin{abstract}
The electron and photon structure functions are compared.
Advantages of the electron structure function are demonstrated. 
At very high momenta probabilistic (partonic) 
interpretation
can be preserved despite strong $\gamma$-$Z$ interference. 
At present energies analyses of both the  electron and the photon 
structure functions give an important test of
 the experimentally applied methods. Predictions for the electron
structure function at present and future momenta are given.
\end{abstract}

\begin{figure}[h]
\centerline{\epsfig{file=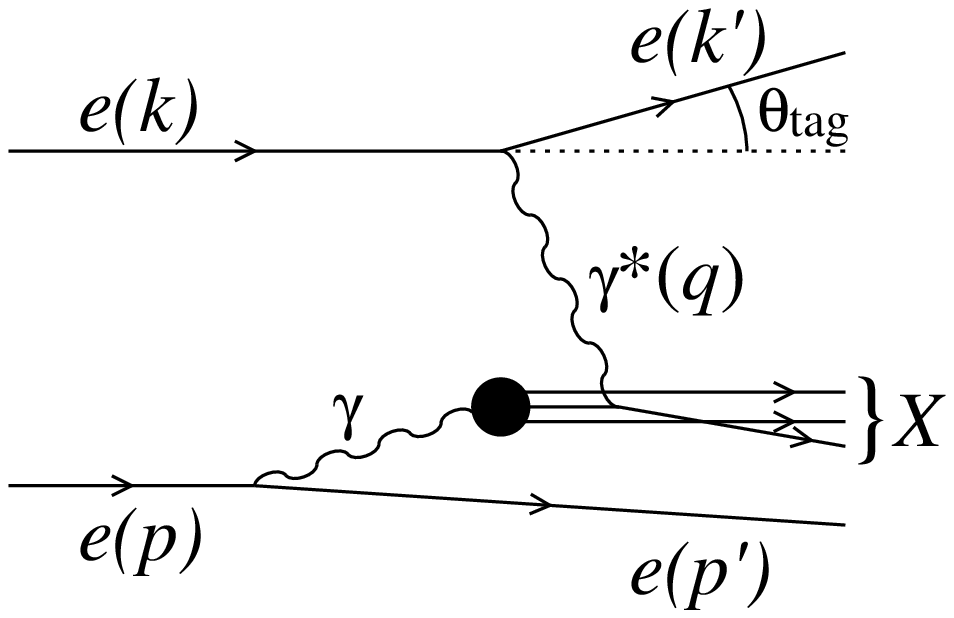,width=5cm}%
\unitlength=1mm
\put(-60,15){\large a)}%
\hspace{20mm}
\epsfig{file=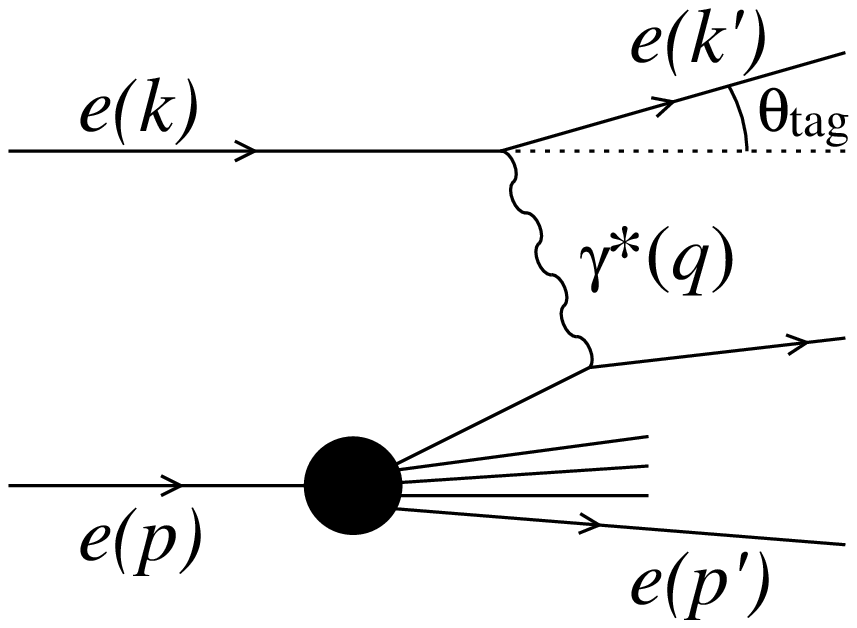,width=5cm}%
\put(-60,15){\large b)}}
\caption{Deep inelastic scattering on a photon (a) and electron (b) target}
\label{f.DIS}
\end{figure}

Let me start with the well known measurement of the photon
 structure function, displayed graphically 
in Fig.~\ref{f.DIS}a. The
tagged (upper) electron emits a probing photon whereas the untagged
(lower) one goes nearly along the beam, emitting the target photon. The
large scale $Q^2=-q^2$, is determined solely by the tagged electron. The anti-tagging 
condition (if present) requires the virtuality of the target photon
 $P_{\gamma}^2$ to be 
between its kinematically allowed minimum and 
 certain fixed maximum $P^2$:
\begin{equation}
-(p-p')^2 \equiv P_\gamma^2 \le P^2 
\,.
\end{equation}

In practice the average photon virtuality is very close to zero 
and the ususal interpretation says 
that we are measuring the real photon structure function.
This photon is clearly not a beam particle and
has the energy diffused 
according to the equivalent photon (Weizs\" acker-Williams \cite{WW})
spectrum $f^e_{\gamma}$. 
The measured cross-section for the production of a hadronic system $X$, expressed in terms of the 
photon structure functions $F_2^{\gamma}$ and 
$F_L^{\gamma}$ reads:
\begin{eqnarray}
\label{fot}
&&{{d^3\sigma _{ee\rightarrow eX}}\over {dzdQ^2}}
= 
{2\pi \alpha^2\over x^2Q^4} 
 [(1+(1-y)^2)F^{\gamma}_2(x,Q^2,P^2)-y^2F_L^{\gamma}(x,Q^2,P^2)]
 f_{\gamma}^e(z/x,P^2)dx
\end{eqnarray}
where 
\begin{equation}
y=1-(E_{\rm tag}/E)\cos ^2(\theta_{\rm tag}/2)
\end{equation}
and
$x$ ($z$)  are fractions of parton  momentum with respect
to the photon (electron).
We assume here, as well as in Eq. (\ref{ele}) below, 
that the probing 
boson is photon only. The Z boson contribution will be included for 
the predictions at high $Q^2$.

Few remarks are important for further considerations. 
First, 
the splitting of the process into a distribution of photons inside electron $f_{\gamma}^e$ 
and that
of partons inside the photon $F_2^{\gamma}$ is an approximation. 
 Second,
in order to fix $x$, one is forced to measure --- in addition to the tagged 
electron --- the hadronic momenta. In fact, 
\begin{equation}
x= {Q^2 \over Q^2+P_\gamma^2 + W^2}
 \approx
 {Q^2 \over Q^2 + W^2}
\,,
\end{equation}
where 
$W$ is the invariant mass of the produced hadronic system $X$. Its determination 
is more difficult than of other (tagged electron) variables.
The uncertainty in  the determination of the
$x$ variable is the source of biggest errors in the analysis. 
The data are indirectly biased by theoretical assumptions. 

Some  of the above problems can be avoided when we introduce the
structure function of the electron (Fig.~\ref{f.DIS}b). To see how it works 
let us first write the cross-section (\ref{fot}), this
time in terms of the electron structure functions $F_2^e$ and $F_L^e$:
\begin{equation}
\label{ele}
{{d^2\sigma _{ee\rightarrow eX}}\over {dzdQ^2}} =
 {2\pi \alpha^2\over z Q^4}
[(1+(1-y)^2)F^e_2(z,Q^2,P^2)-y^2F_L^e(z,Q^2,P^2)].
\end{equation}

No unfolding
procedure is necessary to obtain $F_2^e$, and its argument $z$ --- the parton momentum fraction
with respect to the electron --- is measured, as in the standard deep inelastic 
scattering, by means of the tagged electron
variables only:
\begin{equation}
z={Q^2\over {2pq}}=
{\sin^2(\theta_{\rm tag}/2) \over E/E_{\rm tag} -\cos^2(\theta_{\rm tag}/2) }.
\end{equation}
There is no need {\it a priori} to reconstruct the hadronic mass $W$
(In practice a lower limit cut on $W$ is imposed due to 
difficulties in experimental reconstruction of very low hadronic masses.). 
All these features cause that
the same experiment can produce more precise 
and analysis independent
data when looking at the 
electron structure. 
What is most important
--- the electron 
structure function contains the same information about 
QCD  as the photon one
and is known theoretically with at least the same accuracy. 
Moreover, it allows to avoid  
problems which arise in the photon structure function
at very high energies.

The construction of the QCD electron structure function 
can be summarized in two steps presented in detail in Refs. \cite{ActaESF,QCDesf}.
First we calculate the splitting function of the electron into
a quark/anti-quark. In this process we allow,
in addition to the photon, for the exchange of $Z$ and $W$ bosons even if at 
present energies the photon contribution dominetes (this generalisation
allows for the $\gamma-Z$ interference).
In the second step we construct the $Q^2$-evolution equations for the quark and gluon 
densities inside the electron $q(z,t)$ and $G(z,t)$. 
\begin{figure}
\def\szer{3.1in}
\centerline{\epsfig{file=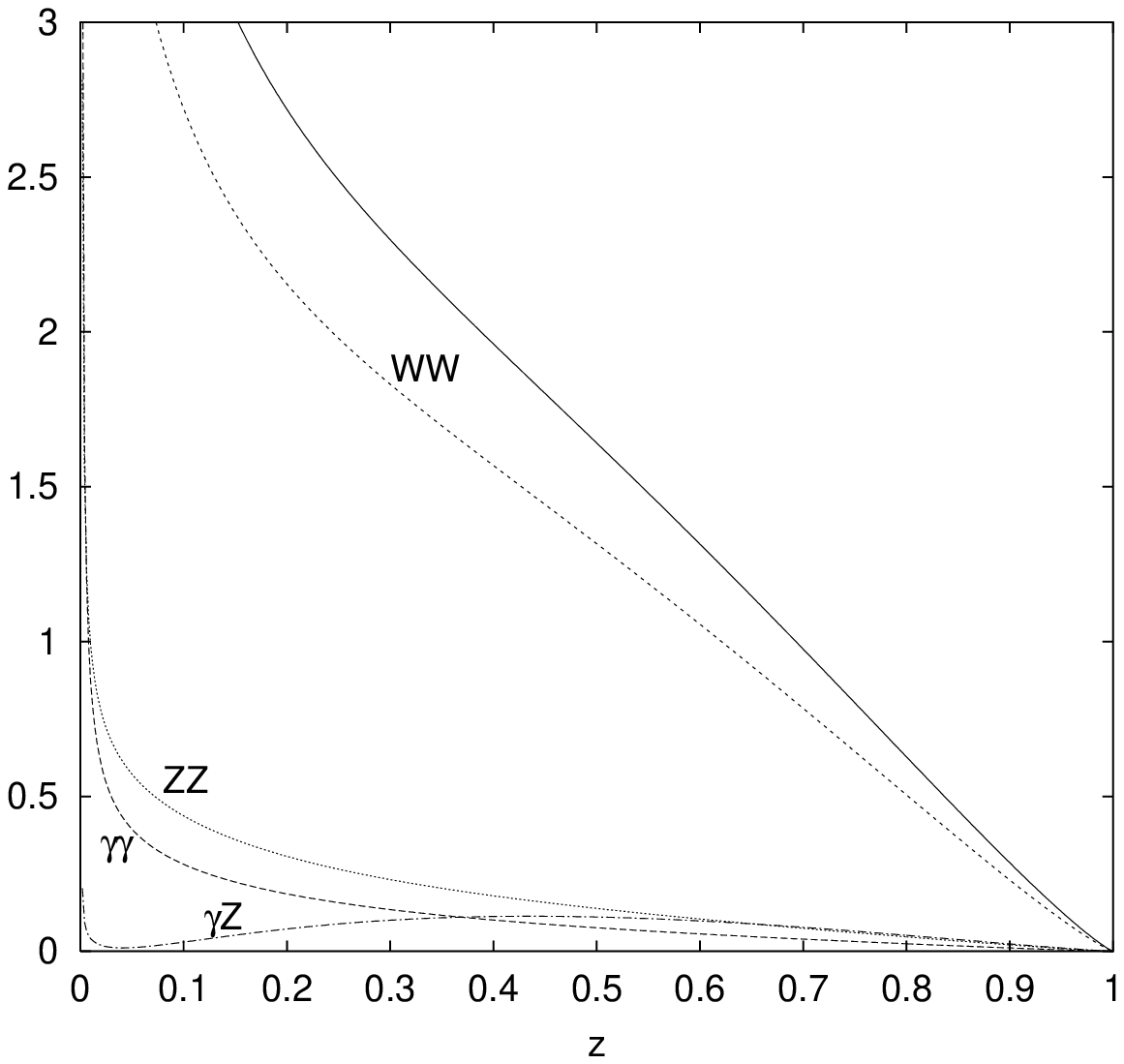,width=\szer}
\unitlength=1mm
\put(-25,65){$z d^{\rm as}(z)$}%
\epsfig{file=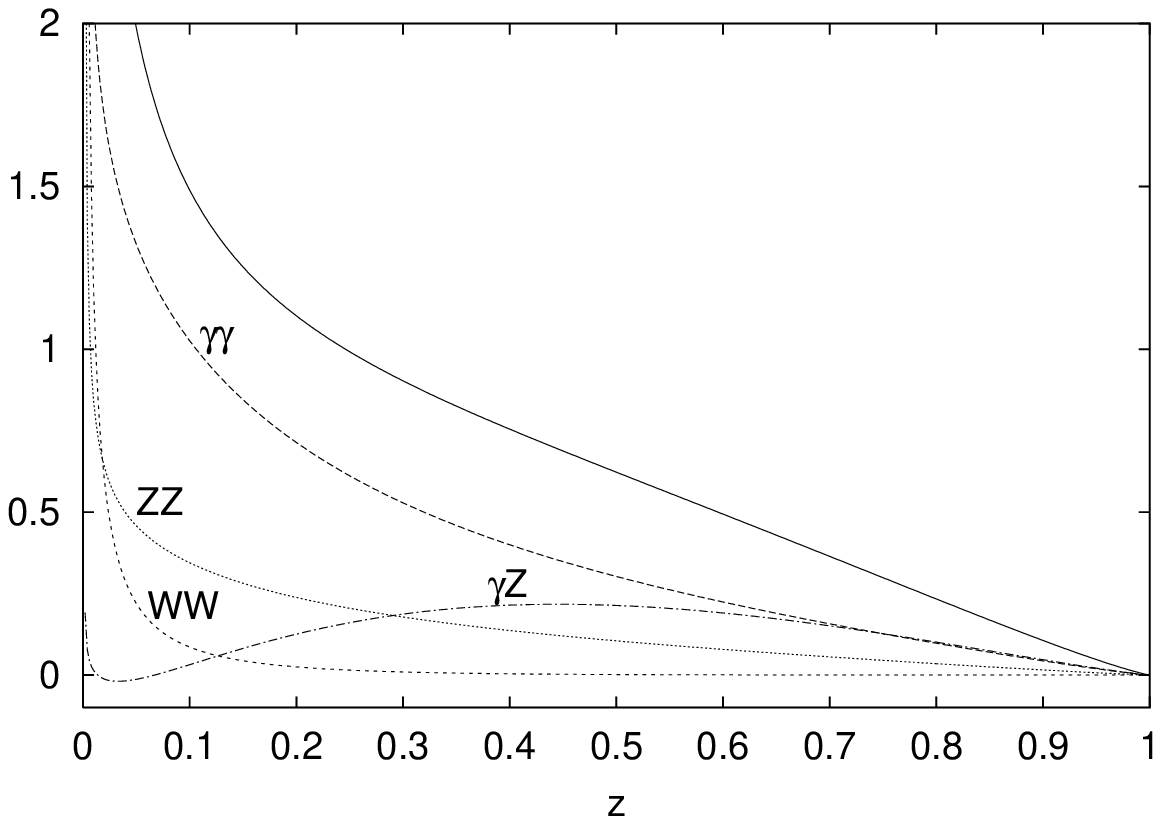,width=\szer}%
\put(-25,45){$z u^{\rm as}(z)$}}
\caption{Asymptotic quark distributions  --- solid line. 
The other lines show different contributions as labeled}
\label{f.epd}
\end{figure}
I skip an interesting discussion of scales entering the process
(see Refs. \cite{ActaESF,QCDesf}). Let me only present the 
asymptotic solutions to the evolution equations in the case with fixed
 antitag momentum squared $P^2$. 
They take the form ($t=log(Q^2/\Lambda_{QCD}^2)$ and
 $t_1=log(P^2/\Lambda_{QCD}^2))$:
\begin{equation}
q(z,t) \simeq
\left( {\alpha \over 2\pi} \right)^2 
q^{\rm as}(z) \,  t_1 t,
\;\;\;\;\;\;
G(z,t) \simeq 
\left( {\alpha \over 2\pi} \right)^2 
G^{\rm as}(z) \,  t_1 t
\label{asdef}
\end{equation}
with $q^{\rm as}(z)$ and $G^{\rm as}(z)$ being given by known, $t$-indepen\-dent
integral equations \cite{ActaESF,QCDesf}.
Their numerical solutions are shown in Fig.~\ref{f.epd}.  As expected 
in the asymptotic 
region all bosons contribute. What is interesting, 
the $\gamma$-$Z$ interference term
enters with strength comparable to the contribution of the $Z$ boson itself.  
This means that 
the notion of separate equivalent bosons breaks down at
very high momenta. The 
electron structure function  takes correctly 
into account interference effects,
preserving at the same time  probabilistic (partonic) interpretation. 
One should keep in mind that only in the asymptotic region  contributions
from all intermediate bosons
entering the splitting function  are comparable. At lower energies
the photon contribution  dominates.

\def\GeV{\,{\rm GeV}}
\begin{figure}[b]
\hbox{\epsfig{file=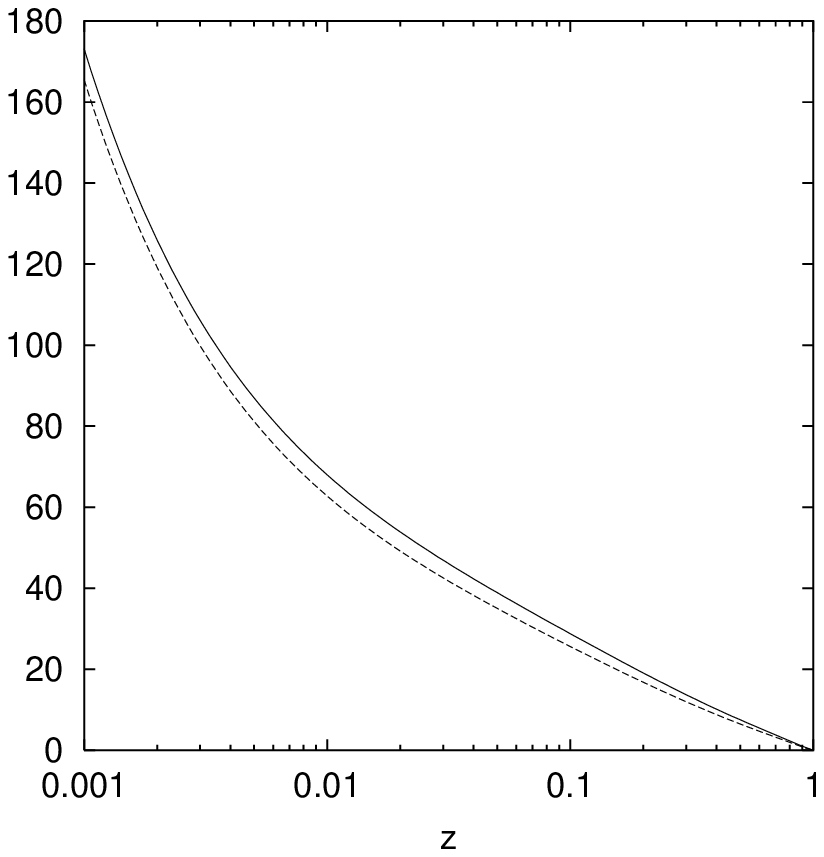,width=3.1in}%
\unitlength=1mm\put(-30,60){$ F_2^e/\alpha^2 $}
\put(-62,15){\Large a)}
\hspace{5mm}
\hbox to 3in {\vbox to 79mm {
\epsfig{file=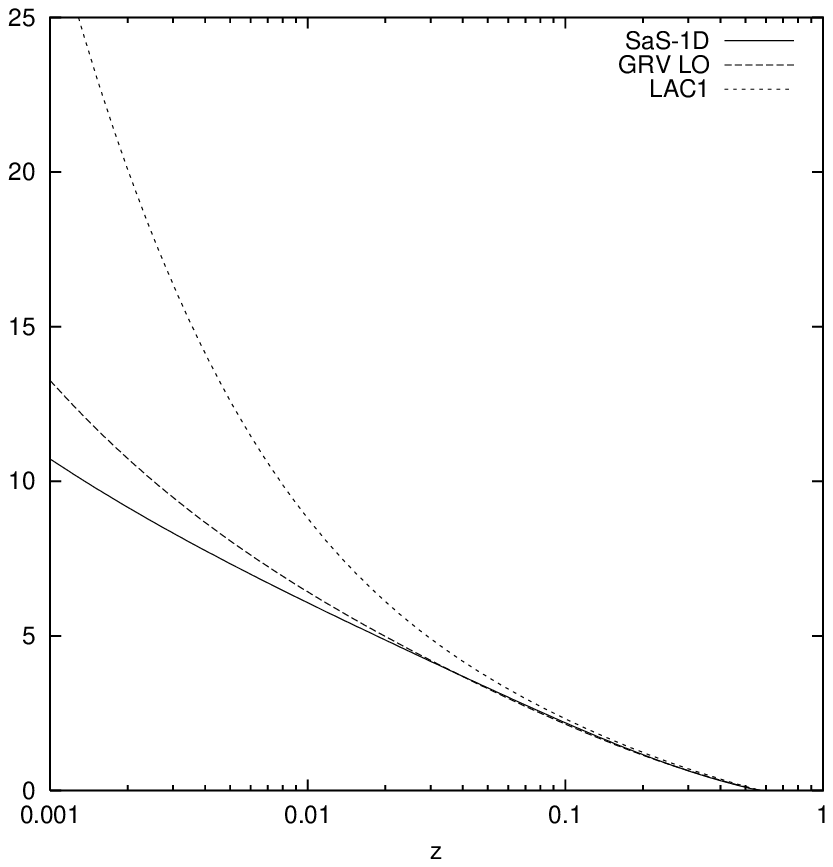,width=3in}\put(-30,60){$ F_2^e/\alpha^2 $}%
\vss}\hss}%
\put(-65,15){\Large b)}
\hfill}
\caption{Predicted value of the electron structure function 
$F^e_2(z)/\alpha^2$:
a) asymptotic solution
at $Q^2=50000\GeV^2$ and 
$P^2$ up to $1000\GeV^2$ (solid line), the contribution 
from the photon only is also shown (broken line);
b) at $Q^2=12\GeV^2$ with the cut on 
the hadronic mass $W>3\GeV$ from different 
parametrizations of the photon structure function:
SaS (solid line), GRV (broken line), LAC1 (dotted line).}
\label{f.F2e}
\end{figure}

Is it possible to measure  the above mentioned new effects in the 
next generation of $e^+e^-$ colliders? 
The answer is given in Fig.~\ref{f.F2e}a
where the electron structure function $F_2^e$ is shown
for  CLIC \cite{CLIC} energies. 
In this figure we use our asymptotic solutions
for the $z$ dependence 
but we multiply each contribution by the appropriate, finite logarithm. 
At such high $Q^2$ 
the electron is probed of course by both the photon and the $Z$ boson.
The effect of $Z$ and $\gamma$-$Z$ 
interference terms in the structure function 
is of the order of 5 to 15\%. Effects of even larger 
 size can be observed in double-tag
experiment \cite{resolec4}. 

At present energies, where the $W$ and $Z$ contributions are negligible, 
one can reanalyze the existing data in terms of the
electron structure function. As already mentioned this can be treated 
also as a consistency check of both photon and electron structure. 
Having a parametrization of the photon structure function which 
describes well the existing data  
we can predict the electron structure function by taking the 
convolution of this parametrization with the equivalent photon spectrum. 
The curves resulting from some popular parametrizations 
\cite{parametrizations} are shown in Fig.~\ref{f.F2e}b.

Let us add a few final remarks. 
First concerns the study of the
virtual photon structure \cite{virt}. The analysis can be reformulated 
in terms of the $P_\gamma^2$ dependence of the electron structure function. 
Studying a real, convention independent object is the first advantage. 
Another one is the fact that
at very high virtualities the $Z$ admixture and the $\gamma$-$Z$
interference are properly taken into account.

We also comment on the QED structure function of the photon.
It is obtained from the process $e^+e^- \rightarrow e^+e^-\mu^+\mu^-$
by dividing out the (approximate) equivalent photon distribution and 
assuming some average photon virtuality.
The use of the QED electron structure function avoids these problems.
The exactly known (in given order of $\alpha$) electron 
structure function can be compared directly with the electron data.

Finally, the photon structure function has been also measured 
\cite{dijets} in di-jet production at HERA. Again the extraction of the $x$
variable is difficult. In addition to jets, one has to measure 
essentially  the whole hadronic system in order to obtain the photon 
energy. The data, when presented in terms of the electron 
structure function, require only measurement of the two jets. 

To summarize, we propose to look at the electron as surrounded by a QCD
cloud of quarks and gluons (in order $\alpha^2$), very much like it is
surrounded by a QED cloud of equivalent photons (in order $\alpha$). We
argue that the use of the electron structure function in electron
induced processes has important advantages over the photon one.
Experimentally it leads to more precise, convention independent data.
Theoretically it allows for more careful treatment of all variables. It
also takes into account all electroweak gauge boson contributions,
including their interference, which will be important in the next
generation of $e^+e^-$ colliders. At present energies it should
certainly be used as a cross-check of the photon structure analysis.
In fact two experimental groups at CERN are currently performing 
the electron structure function analyses.
 

\end{document}